\documentclass[preliminary,copyright,creativecommons]{eptcs}
\providecommand{\event}{WORDS 2011} 
\usepackage{breakurl}      

\usepackage[dvips]{graphicx}
\usepackage{amsmath}
\usepackage{psfrag}
\usepackage{amssymb}
\usepackage{epsf}
\usepackage{epsfig}
\usepackage{color}

\usepackage{tikzexternal}
\newcommand{\tr}{\vphantom{A}^t}

\newcommand{\N}{\mathbb{N}}

\title{Uniformly balanced words    with linear complexity and prescribed letter frequencies }
 
\author{Val\'erie Berth\'e
\institute{Laboratoire d'Informatique Algorithmique : Fondements et Applications\\
 Universit\'e Paris Diderot\\
Paris 7 - Case 7014\\
F-75205 Paris Cedex 13, France }
\email{berthe@liafa.jussieu.fr}
\and
S\'ebastien  Labb\'e
\institute{Laboratoire de Combinatoire et d'Informatique Math\'ematique,\\
Universit\'e du Qu\'ebec \`a Montr\'eal,\\
C.P. 8888 Succursale ``Centre-Ville'', Montr\'eal (QC), Canada H3C 3P8}
\email{labbe.sebastien@courrier.uqam.ca}}
\def\titlerunning{Uniformly balanced words}
\def\authorrunning{V. Berth\'e,  S. Labb\'e}
\date{ }

\begin{document}
\maketitle

\begin{abstract}
We consider    the following problem.
Let us fix a  finite alphabet ${\mathcal A}=\{1,2,\cdots,d\}$;  for  any   $d$-uple of  letter frequencies $(f_1,\cdots,f_d) \in [0,1]^d$
with $\sum_{i=1}^d f_i =1$,  
how to construct   an infinite word  $u$  over the  alphabet  ${\mathcal A}$ satisfying
the following conditions:
$u$  has  linear complexity function,
$u$   is uniformly balanced,
the   letter frequencies in $u$ are given by  $(f_1,\cdots,f_d) $.
This   paper investigates   a  construction method for such words
based on the use of  mixed   multidimensional continued fraction algorithms.
\end{abstract}
\bigskip



\section{Introduction}\label{sec:intro}

We consider    the following problem:
let us fix a  finite alphabet ${\mathcal A}=\{1,2,\cdots,d\}$;  for  any   $d$-uple of  letter frequencies $(f_1,\cdots,f_d) \in [0,1]^d$
with $\sum_{i=1}^d f_i =1$,  
how to construct   an infinite word  $u$  over the  alphabet  ${\mathcal A}$ satisfying
the following conditions:
\begin{enumerate}
\item
$u$  has  linear complexity function
\item 
$u$   is uniformly balanced
\item
the   letter frequencies in $u$ are given by  $(f_1,\cdots,f_d) $.
\end{enumerate}
Let us first  recall several  definitions in order to  clarify the previous statement.
A word  $u \in {\mathcal A}^{\mathbb{N}}$ is said to be  {\em uniformly balanced} if there exists    a constant $C>0$ such  that
for any pair   of factors  of the same  length $v,w$ of $u$, and for any   letter $i \in  \mathcal{A}$,
$$||v|_i - |w|_i| \leq C,$$
where the notation  $|x|_j$ stands for the  number  of  occurrences  of the  letter $j$ in the factor $x$.
A  word  $u$ has {\em linear complexity function } if there exists a constant $C'>0$ such  that
the  number of factors of $u$ of length $n$  is  smaller  than $C' \cdot n$, for every positive  integer $n$. The {\em frequency}  $f_i$ of a  letter $i \in {\mathcal A}$
in  $u=(u_n)_{n \in \mathbb{N}}$ is defined 
as the limit (when $n$ tends towards infinity), if it exists, of 
the number of occurrences
of  $i$  in $u_0 u_1 \dots u_{n-1}$   divided by $n$.

This problem   has  several motivations.   The first one  comes  from  discrete geometry:    such  an infinite word
can be seen as a coding  of a discrete  line in $\mathbb{Z}^d$. Indeed  one associates with      any infinite word
over the alphabet  ${\mathcal A}$ a  broken line     obtained   as a  stair made of a  union of segments of unit length  directed  according to the  coordinate axes, 
 whose  vertices  are obtained    by replacing each of the letters   of $u$  by one of the      canonical  basis    vectors
 and by concatenating these vectors.  Let 
  ${\mathbf l}:A^*\to\N^n, \ w\mapsto \tr(|w|_{a_1},\ldots,|w|_{a_n})$
    stand for  the {\em   abelianisation map}   or 
   the {\em Parikh mapping}.
   More precisely, the set of vertices  of  this broken line    is  equal to 
 $\{{\mathbf l}(u_0 \cdots u_{N-1}) \mid
N\in \N\}$. The question is to know how  well the line  associated with the word $u$  approximates  the   Euclidean line
  directed by the  vector  of letter  frequencies of $u$,  when they exist.    There exist   various    strategies  for defining and generating discrete lines in the three-dimensional space. With no claim
   for being exhaustive, let us quote  e.g.
\cite{andres,BB09,FigRev96,toutant_characterization_2006}. 
Nevertheless, they    do not   fulfill     Condition  1.  on the linear complexity. Note that the  notion of discrete line  defined  in \cite{andres} corresponds
to billiard  words.      Condition 1. means here   that  these  discrete lines  are  ``simple'' in terms of number of local configurations.

The second motivation comes from   symbolic dynamical systems and  Diophantine approximation:
is it possible  to define a Rauzy fractal     associated  with any   translation  of the  torus?
More precisely,   assume   we are given a      translation $ x \mapsto   x+ (\alpha_1, \cdots  \alpha_d)$   defined on $
\mathbb{T}^d = {\mathbb R}^d / {\mathbb Z}^d$;     the  Rauzy fractal ${\mathcal R}$ associated  with   an infinite  word $u $  over  the $d$-letter alphabet ${\mathcal A}$ is defined  
  by  projecting    along  the  frequency vector of $u$   on a transverse   hyperplane  the vertices   of the broken line associated with 
$u$  (such as described   above)  and   then, by  taking the closure. For more on Rauzy fractals, see  e.g.  \cite{BertheSiegel2005}. The problem now becomes the following:
   is it    possible  to construct an infinite  word $u $  over  the $d$-letter alphabet ${\mathcal A}$
 such that
${\mathcal R}$ is a  compact set that tiles periodically this transverse hyperplane and such that
$u$ has linear complexity? Let us explain in this context the  requirement  concerning linear  complexity (Condition 1.):  we would like to recover from  the
dynamical and combinatorial properties of the  infinite  word $u$  arithmetical information   on the parameters  underlying the  translation on the torus.
This will be easier if   $u$   has low complexity function,  i.e., a low  numbers of factors.  
Let us quote as a further motivation uniform distribution and   the  so-called   chairman assignment problem, see  e.g.   \cite{tij},  and the references  therein.

There exist families of words  that   satisfy  Conditions 2. and  3.    but not Condition 1. Billiard words   are defined as codings of  trajectories of billiards in a cube;
they  are shown to have quadratic complexity  (see   \cite{amst1,Baryshnikov:1995}).  They satisfy  Conditions 2. and  3. Let us also  quote   the construction   described in  \cite{Chevallier:09}   which produces step by step a broken line whose vertices  belong to
  $\mathbb{Z}^3$ that approximates a given  direction  by choosing at each step the  closest point. It is proved
  in  \cite{Chevallier:09} that such a  broken line can be  obtained by selecting  integer points   by shifting a polygonal
  window  along the line.  The complexity is here again quadratic.      The corresponding infinite words  satisfy  Conditions 2. and  3. 
  Note that  $1$-balanced words over a higher-alphabet do not seem to be   good candidates
for   describing   discrete segments in the space: not all frequencies can be reached.  Fraenkel's conjecture states  that    the  possible  frequencies   for  $1$-balanced  words   are rational  and uniquely determined, when  they are assumed   to be   distinct \cite{Fra73}. In particular,
when $k=3$, the only possible  $1$-balanced  word is $(1213121)^{\infty}$ (if frequencies are distinct),
up to a permutation of letters  and up to shifts.  For the irrational case, see \cite{Hub00} and \cite{Graham}.  For more references on the subject, see also   the survey \cite{Vui}.
Note also  that Arnoux-Rauzy    words  (see e.g. \cite{arnrau91,CFZ,CFM})  are 
infinite words   that do not satisfy Condition  2.,  such as proved  in  \cite{CFZ}, but that do  satisfy Conditions 1. and 3.  Furthermore,
they are not defined for every $d$-uple of letter frequencies, but     only for a set of zero measure in $[0,1]^d$.  For an illustration, see Figure \ref{fig:AR}.
\section{Multidimensional continued fractions and frequencies}
  
  The strategy we consider here     for   constructing  infinite words satisfying  the three  above mentioned conditions  consists in   applying a multidimensional continued fraction algorithm to the
  frequency vector $(f_1, \cdots,f_d)$, according to \cite{BLab}.  We then   associate  with the steps of the algorithm substitutions, that is, rules that  replace 
letters by words,   with these substitutions having   the  matrices   produced by the continued fraction algorithm   as incidence matrices. More precisely,
 a {\em substitution} $\sigma$ over the   alphabet~${\cal A}$ is an endomorphism of the free monoid
${\cal A}^*$, and the  {\em incidence matrix}  of the substitution $\sigma$ is the square matrix $M_{\sigma} $ with entries  $m_{i,j} = | \sigma(j)|_i$ for all $i,j  \in {\mathcal A}$.

Let us  recall  the most classical multidimensional continued fraction algorithms      such    as  described e.g.  in \cite{schweiger}, and in \cite{CFZ,CFM,WZ} for Arnoux-Rauzy  algorithm.
For the sake of simplicity, we express them  in dimension $d=3$:
\begin{itemize}

\item   {Jacobi-Perron:}  
let 
$0\leq u_1,u_2 \leq u_3$,
$$(u_1,u_2,u_3) \mapsto (u_2- [\frac{u_2}{u_1}] u_1, u_3- [\frac{u_3}{u_1}] u_1, u_1).$$
\item   {Brun: }
 we subtract the second largest entry from the largest one;   for instance, if  
   $0 \leq u_1\leq u_2 \leq u_3$,  
$$(u_1,u_2,u_3) \mapsto  (u_1, u_2, u_3 -u_2).$$

\item   {Poincar\'e:}  we subtract  the second largest entry to the largest one,  and   the smallest  entry  from the  second largest one; for instance, if 
$0 \leq u_1\leq u_2 \leq u_3$,
$$(u_1,u_2,u_3) \mapsto  (u_1, u_2-u_1, u_3 -u_2).$$

\item   {Selmer: }
 we subtract the smallest  positive entry   from the largest  one;   for instance, if   $0  < u_1\leq u_2 \leq u_3$,
$$(u_1,u_2,u_3) \mapsto  (u_1, u_2, u_3 -u_1).$$

\item   {Fully subtractive: }
we subtract the smallest   positive entry   from all  the largest  ones;   for instance, if 
  $0  <  u_1\leq u_2 \leq u_3$,
$$(u_1,u_2,u_3) \mapsto  (u_1, u_2-u_1, u_3 -u_1).$$

\item {Arnoux-Rauzy:}
let  $0 \leq u_1\leq u_2 \leq u_3$  with  $u_3 \geq u_1+u_2$,
$$(u_1,u_2,u_3) \mapsto  (u_1, u_2, u_3 -u_1 -u_2).$$
otherwise the algorithm stops.

\end{itemize}
Let $T$  be one of these   algorithms applied to  some vector $(f_1,f_2,f_3) \in [0,1]^3$.
With each matrix $M$ produced by  $T$, we associate  a substitution whose incidence matrix
is given by $M$.   We thus obtain a  word  by iterating these   substitutions in an $S$-adic way.
 We recall that a word  is said to be {\em  $S$-adic}  if it is generated by     composing   a  finite number of substitutions. This covers     various  families of  words
 with a rich   dynamical  behavior such as Sturmian sequences;  for more on $S$-adic words, see  e.g.  \cite{ABFogg,Durand}.



 

\section{Fusion algorithms}

We can also mix these   algorithms by  performing at each step  one  among  these   rules, and this still yields
$S$-adic generated words. We call such  algorithms  {\em fusion algorithms}. We  focus on 
fusion algorithms obtained by  applying Arnoux-Rauzy algorithm when possible, and  otherwise,    consistently  one  algorithm   among    Brun, Poincar\'e, Selmer,
or the Fully Subtractive algorithms. Indeed, experimental studies   indicate that a combination of  Arnoux-Rauzy steps with  Brun steps,  or   with Poincar\'e  steps 
produces     good   performances   (see Table~\ref{table_tijdeman} and    Figure  \ref{fig:PAR} below), and
even better performances    than   when performing only one  algorithm.     Furthermore,  this allows us
to   exploit and extend  the good mean  behaviour of Arnoux-Rauzy  algorithm   to a larger set of parameters  (compare Figure \ref{fig:AR} and Figure \ref{fig:PAR}).

The aim of this lecture is to  study   the properties of     such  fusion   algorithms for both  finite  (rational frequencies) and infinite expansions (irrational  frequencies).
In particular, we will  focus on 
the almost everywhere  convergence properties    and  ergodic  properties of    these fusion  algorithms  when the   frequency  vector
has irrational coordinates. The  proof relies on classic techniques
   such as described  e.g. in \cite{schweiger}.

\begin{table}[h]
\begin{tabular}{c|c|c|c|c}
 & Minimum & Mean & Maximum & Std \\ 
\hline
Arnoux-Rauzy & 0.6000 & 0.9055 & 1.200 & 0.1006 \\ 
\hline
Fully subtractive & 0.6000 & 5.982 & 13.92 & 4.388 \\ 
\hline
Fully subtractive as possible & 0.6000 & 4.172 & 25.00 & 4.440 \\ 
\hline
Selmer & 0.5000 & 2.184 & 12.75 & 2.070 \\ 
\hline
Brun & 0.5000 & 1.114 & 2.000 & 0.2664 \\ 
\hline
Brun Multiplicative & 0.6000 & 1.117 & 2.000 & 0.2681 \\ 
\hline
Poincar\'e & 0.6000 & 2.527 & 11.13 & 2.261 \\ 
\hline
Jacobi-Perron & 0.6000 & 2.731 & 25.00 & 3.456 \\ 
\hline
Random reduction & 0.5000 & 2.426 & 24.99 & 2.779 \\ 
\hline
Fusion of Arnoux-Rauzy and Fully subtractive & 0.6000 & 1.095 & 2.800 & 0.3105 \\ 
\hline
Fusion of Arnoux-Rauzy and Selmer & 0.6000 & 0.9678 & 1.450 & 0.1438 \\ 
\hline
Fusion of Arnoux-Rauzy and Brun Multiplicative & 0.6000 & 0.9132 & 1.400 & 0.1143 \\ 
\hline
Fusion of Arnoux-Rauzy and Poincar\'e & 0.6000 & 0.8941 & 1.200 & 0.09733 \\ 
\end{tabular}

\caption{Statistics (minimum, mean, maximum, standard deviation) for  the  discrepancy    for triplets of  nonnegative rational  vectors   $(a_1/N,a_2/N,a_3/N)$ such that $a_1+a_2+a_3=N$ with  $N=100$.}
\label{table_tijdeman}
\end{table}


Consider now the case of  rational frequencies.  Table~\ref{table_tijdeman} displays  some experimental results.  We work here in dimension $d=3$
with  rational  frequency vectors of the form ${\bf f}=(a_1/N,a_2/N,a_3/N)$, with $a_i \in  \mathbb{N}$, $i=1,2,3$, and with  $a_1+a_2+a_3=N $ being a positive integer.  
We apply  a  fusion algorithm  to  such a triplet, until we reach   a vector whose entries are all equal to $0$ but one. This produces a  finite sequence of matrices, and thus, of   substitutions, having these  matrices  as  incidence matrices. Note that we have several choices for these substitutions, even  if 
the incidence matrices have  entries in $\{0,1\}$.  Given a matrix $M$, we  thus have to  decide in which order  letters will be   chosen in the image of a  letter by  a substitution $\sigma$ having 
$M$ as incidence matrix. We choose  as a convention    to  put  the most frequent  letter first.  (This  (partly) explains why the triangles  obtained in Figure  \ref{fig:P20}, \ref{fig:FS},  \ref{fig:P}, \ref{fig:AR}, \ref{fig:PAR} are not  perfectly symmetric.)
Let us apply  now to ${\bf f}$ a finite  sequence of   steps  of a fusion algorithm 
together with a choice  of  substitutions associated with the produced matrices. One has
$
{\bf f }= M_1 \cdots M_n {\bf f }_n
$,
where  the vector ${\bf f}_n $  has  two coordinates equal to~$0$, and  one non-zero coordinate of index, say 
$w_n \in \{1,2,3\}$. The associated   substitutions  are denoted by $\sigma_k$, for $1 \leq k \leq n$.  The following  diagram  illustrates how we produce
finite words $w$ with frequency  vector  ${\bf f}$:


\begin{center}
\begin{tikzpicture}[descr/.style={fill=white,inner sep=2.5pt}] 
\matrix (m) [matrix of math nodes, row sep=1.5em, column sep=3em, text height=1.5ex, text depth=0.25ex] 
 {{\bf f} = {\bf f}_0 & {\bf f}_1 & {\bf f}_2 & \cdots & {\bf f }_n \\ 
    w= w_0  &w_1 & w_2 & \cdots & w_n \in \{1,2,3\}\\ };
\path[<-,font=\scriptsize] 
(m-1-1) edge node[above] {$M_1$} (m-1-2) 
(m-1-2) edge node[above] {$M_2$} (m-1-3) 
(m-1-3) edge node[above] {$M_3$} (m-1-4) 
(m-1-4) edge node[above] {$M_n$} (m-1-5) ;
\path[<-,font=\scriptsize] 
(m-2-1) edge node[above] {$\sigma_1$} (m-2-2) 
(m-2-2) edge node[above] {$\sigma_2$} (m-2-3) 
(m-2-3) edge node[above] {$\sigma_3$} (m-2-4) 
(m-2-4) edge node[above] {$\sigma_n$} (m-2-5) ;
\end{tikzpicture} 
\end{center}  

The experimental results  of Table~\ref{table_tijdeman} indicate that the     fusion algorithm obtained when applying Arnoux-Rauzy algorithm when possible, and  otherwise,       Poincar\'e algorithm, behaves in an efficient way  with respect to   the discrepancy. The  {\em discrepancy }    of  a  finite word  $u_0\cdots u_n \in {\mathcal A}^{n+1}$ is defined  as 
 $$\max_{i \in {\mathcal A}, \ 0 \leq k \leq n}    | f_i  \cdot k - | u_0\cdots u_k| _i |.$$  This distance is considered e.g.  in \cite{tij} and \cite{Adam},
 and is intimately connected with the following   balance measure.  The  {\em balance} of $u_0\cdots u_n \in {\mathcal A}^{n+1}$ is defined as 
$$\max_{i \in {\mathcal A}, \  |v|=|w|}  ||v|_i - |w|_i| ,$$ 
(here $v,w$ are     factors   of $u$ of the same length $|v|=|w|$).  We have chosen here to use the discrepancy for our  numerical experiments in order
to compare our results with the  bound  discussed  in  \cite{tij}.
Indeed, in   \cite{tij},   an  algorithm   is given that produces, for any given  frequency vector $(f_1, \cdots, f_d)$,   an infinite word   whose
 discrepancy   is smaller than or equal to $1-1/(2d-2)$ (this yields $3/4$ for $d=3$). However,   the  lowest possible   asymptotic order  for the  factor complexity of   such a  word     does not seem to be   known; nothing seems  a priori to prevent it  from  being linear.  In the fusion algorithm obtained  by combining Arnoux-Rauzy algorithm with Poincar\'e algorithm,
 one obtains   a mean  discrepancy   equal to $0.8910$ when $N=100$. More generally, Figure  \ref{fig:P20}, \ref{fig:FS},  \ref{fig:P}, \ref{fig:AR}, \ref{fig:PAR} below  illustrate  the behaviour  of the discrepancy 
for   triplets of nonnegative rational  vectors   $(a_1/N,a_2/N,a_3/N)$ such that $a_1+a_2+a_3=N$  for a  given $N$.

\begin{center}
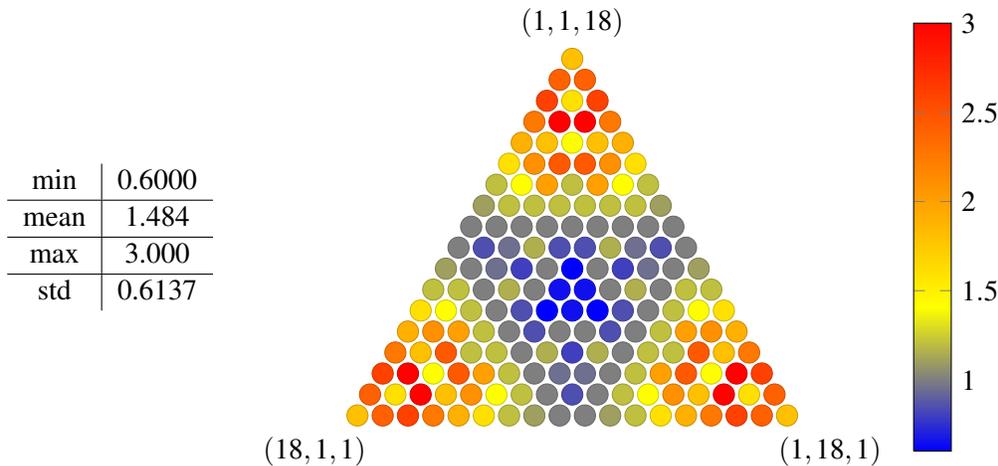
\begin{figure}[ht!]
\begin{minipage}[c]{0.2\linewidth}
\begin{tabular}{c|c}
min & 0.6000 \\
\hline
mean & 1.484 \\
\hline
max & 3.000 \\
\hline
std & 0.6137 \\
\end{tabular}

\end{minipage}
\begin{minipage}[c]{0.8\linewidth}
\begin{tikzpicture}[mark size=4,every mark/.append style={line width=0pt}]
\begin{axis} [clip=false,axis x line=none, axis y line=none, colorbar, 
]
\addplot+[scatter,only marks,scatter src=explicit] 
table[x=xproj,y=yproj,meta=stat] 
{data/table_distance_tijdeman_sum20_oMFF_poincare.dat}; 
\node[] at (axis description cs:0,0) {$(18, 1, 1)$};
\node[] at (axis description cs:1,0)  {$(1, 18, 1)$};
\node[] at (axis description cs:0.5,1)  {$(1, 1, 18)$};
\end{axis} 
\end{tikzpicture} 
\end{minipage}
 \caption{Discrepancy for triplets of nonnegative rational  vectors   $(a_1/N,a_2/N,a_3/N)$ such that $a_1+a_2+a_3=N$ with $N=20$ using Poincar\'e algorithm.}
\label{fig:P20}
\end{figure}
\end{center}

\begin{center}
\begin{figure}[ht!]
\begin{minipage}[c]{0.2\linewidth}
\begin{tabular}{c|c}
min & 0.6000 \\
\hline
mean & 5.982 \\
\hline
max & 13.92 \\
\hline
std & 4.388 \\
\end{tabular}

\end{minipage}
\begin{minipage}[c]{0.8\linewidth}
\begin{tikzpicture}[mark size=0.7,every mark/.append style={line width=0pt}]
\begin{axis} [clip=false,axis x line=none, axis y line=none,colorbar]
\addplot+[scatter,only marks,scatter src=explicit] 
table[x=xproj,y=yproj,meta=stat] 
{data/table_distance_tijdeman_sum100_oMFF_fully.dat}; 
\node[] at (axis description cs:0,0) {$(98, 1, 1)$};
\node[] at (axis description cs:1,0)  {$(1, 98, 1)$};
\node[] at (axis description cs:0.5,1)  {$(1, 1, 98)$};
\end{axis} 
\end{tikzpicture} 
\end{minipage}
 \caption{Discrepancy for triplets of nonnegative rational  vectors   $(a_1/N,a_2/N,a_3/N)$ such that $a_1+a_2+a_3=N$ with $N=100$ using Fully subtractive algorithm.}
\label{fig:FS}
\end{figure}
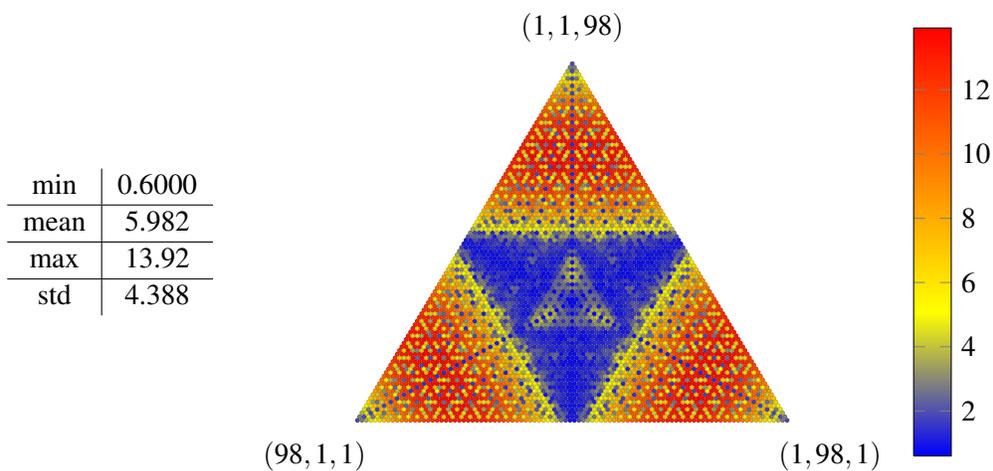
\end{center}

\begin{center}
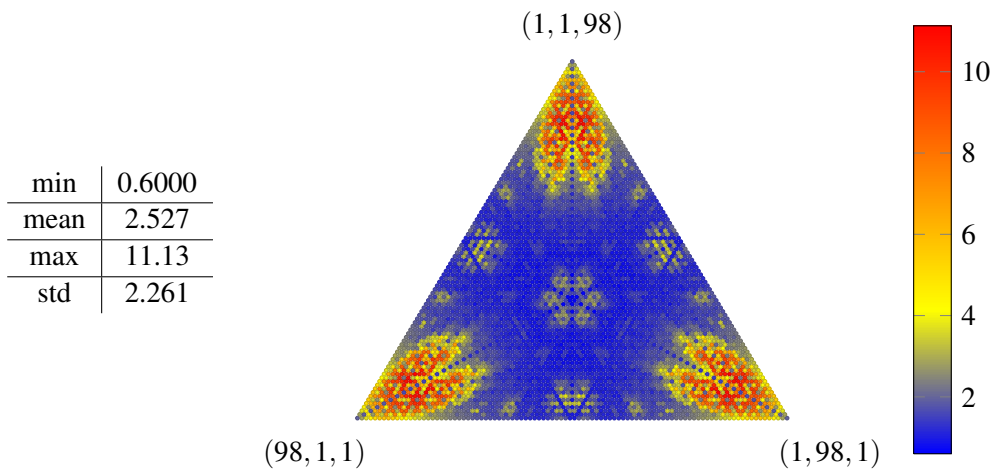
\begin{figure}[ht!]
\begin{minipage}[c]{0.2\linewidth}
\begin{tabular}{c|c}
min & 0.6000 \\
\hline
mean & 2.527 \\
\hline
max & 11.13 \\
\hline
std & 2.261 \\
\end{tabular}

\end{minipage}
\begin{minipage}[c]{0.8\linewidth}
\begin{tikzpicture}[mark size=0.7,every mark/.append style={line width=0pt}]
\begin{axis} [clip=false,axis x line=none, axis y line=none,colorbar]
\addplot+[scatter,only marks,scatter src=explicit] 
table[x=xproj,y=yproj,meta=stat] 
{data/table_distance_tijdeman_sum100_oMFF_poincare.dat}; 
\node[] at (axis description cs:0,0) {$(98, 1, 1)$};
\node[] at (axis description cs:1,0)  {$(1, 98, 1)$};
\node[] at (axis description cs:0.5,1)  {$(1, 1, 98)$};
\end{axis} 
\end{tikzpicture} 
\end{minipage}
 \caption{Discrepancy for triplets of nonnegative rational  vectors   $(a_1/N,a_2/N,a_3/N)$ such that $a_1+a_2+a_3=N$ with $N=100$ using Poincar\'e algorithm.}
\label{fig:P}
\end{figure}
\end{center}

\begin{center}
\begin{figure}[ht!]
\begin{minipage}[c]{0.2\linewidth}
\begin{tabular}{c|c}
min & 0.6000 \\
\hline
mean & 0.9055 \\
\hline
max & 1.200 \\
\hline
std & 0.1006 \\
\end{tabular}

\end{minipage}
\begin{minipage}[c]{0.8\linewidth}
\begin{tikzpicture}[mark size=0.7,every mark/.append style={line width=0pt}]
\begin{axis} [clip=false,axis x line=none, axis y line=none,colorbar]
\addplot+[scatter,only marks,scatter src=explicit] 
table[x=xproj,y=yproj,meta=stat] 
{data/table_distance_tijdeman_sum100_oMFF_ar.dat}; 
\node[] at (axis description cs:0,0) {$(98, 1, 1)$};
\node[] at (axis description cs:1,0)  {$(1, 98, 1)$};
\node[] at (axis description cs:0.5,1)  {$(1, 1, 98)$};
\end{axis} 
\end{tikzpicture} 
\end{minipage}
 \caption{Discrepancy for triplets of nonnegative rational  vectors   $(a_1/N,a_2/N,a_3/N)$ such that $a_1+a_2+a_3=N$ with $N=100$ using Arnoux-Rauzy algorithm. This algorithm is defined only for vectors whose largest entry is greater than or equal to the sum of the other two.}
\label{fig:AR}
\end{figure}
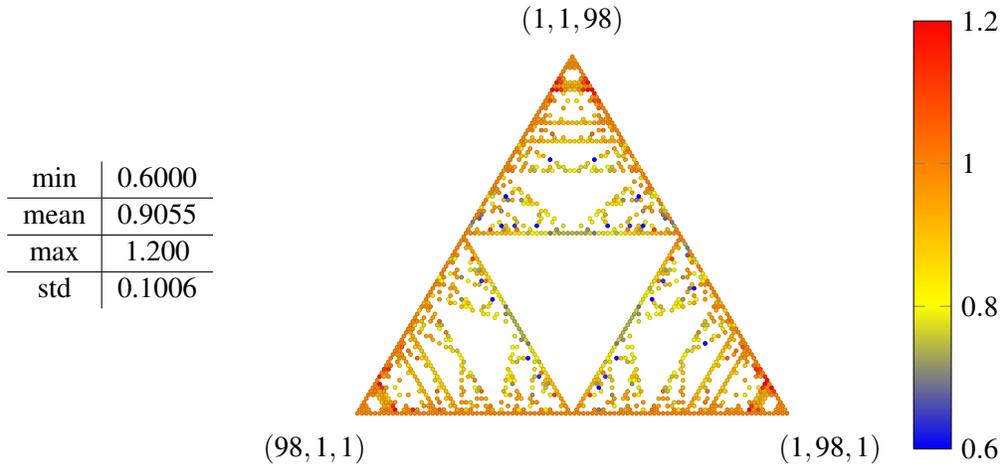
\end{center}

\begin{center}
\begin{figure}[ht!]
\begin{minipage}[c]{0.2\linewidth}
\begin{tabular}{c|c}
min & 0.6000 \\
\hline
mean & 0.8941 \\
\hline
max & 1.200 \\
\hline
std & 0.09733 \\
\end{tabular}

\end{minipage}
\begin{minipage}[c]{0.8\linewidth}
\begin{tikzpicture}[mark size=0.7,every mark/.append style={line width=0pt}]
\begin{axis} [clip=false,axis x line=none, axis y line=none,colorbar]
\addplot+[scatter,only marks,scatter src=explicit] 
table[x=xproj,y=yproj,meta=stat] 
{data/table_distance_tijdeman_sum100_oMFF_arnoux_poincare.dat}; 
\node[] at (axis description cs:0,0) {$(98, 1, 1)$};
\node[] at (axis description cs:1,0)  {$(1, 98, 1)$};
\node[] at (axis description cs:0.5,1)  {$(1, 1, 98)$};
\end{axis} 
\end{tikzpicture} 
\end{minipage}
 \caption{Discrepancy for triplets of nonnegative rational  vectors   $(a_1/N,a_2/N,a_3/N)$ such that $a_1+a_2+a_3=N$ with $N=100$ using a fusion of  Arnoux-Rauzy and Poincar\'e algorithms.}
\label{fig:PAR}
\end{figure}
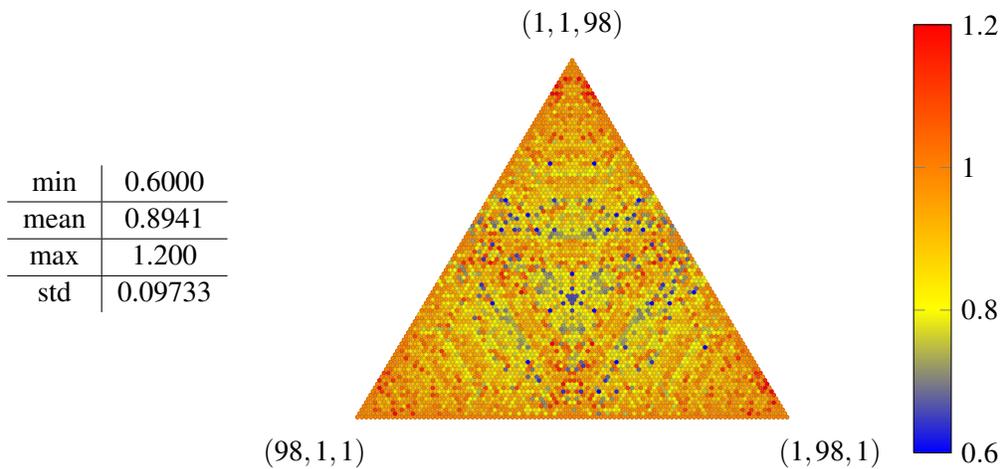
\end{center}

\newpage
\subsection*{Acknowledgements}
We  would  like to thank  warmly J. Shallit for  pointing  out    reference \cite{tij}, as well as R. Tijdeman for useful comments.
This research was driven by computer exploration using the open-source
mathematical software \emph{Sage}~\cite{sage} and its library on Combinatorics on words developed by the \emph{Sage-Combinat}
community, and in particular by the active developers:  A. Blondin Mass\'e, V. Delecroix, S. Labb\'e, T. Monteil and  F. Saliola.

\nocite{*}
\bibliographystyle{eptcs}
\bibliography{Words_Berthe_Labbe.bib}
\end{document}